\documentclass[twocolumn,showpacs,superscriptaddress,amssymb]{revtex4-1}
\date{\today}
\usepackage{graphicx}
\usepackage{multirow}
\usepackage{longtable}
\begin{document}
\title{Effects of formation properties in one-proton radioactivity}

\author{Chong Qi}
\email{chongq@kth.se}
\affiliation{KTH, Alba Nova University Center, SE-10691 Stockholm, Sweden}
\author{Doru S. Delion}
\affiliation{``Horia Hulubei" National Institute of Physics and Nuclear Engineering, 
407 Atomistilor, RO-077125 Bucharest-Magurele, Romania}
\affiliation{Academy of Romanian Scientists, 54 Splaiul Independentei, RO-050085 Bucharest, Romania}
\author{Roberto J. Liotta}
\affiliation{KTH, Alba Nova University Center, SE-10691 Stockholm, Sweden}
\author{Ramon Wyss}
\affiliation{KTH, Alba Nova University Center, SE-10691 Stockholm, Sweden}

\date{\today }

\begin{abstract}
It is shown that the proton formation probability, extracted from
experimental data corresponding to one-proton radioactivity, is divided into
two regions when plotted as a function of an universal parameter. This parameter is derived from a microscopic description of the decay process.
In this way we explain the systematics of proton emission half-lives.
At the same time the formation probability is shown to be a useful quantity to determine
the deformation property of the mother nucleus.
\end{abstract}


\maketitle

Understanding how nuclear many-body systems can self-organize in 
simple and regular patterns is a long-standing challenge in modern 
physics. The first case where this was realized was in  $\alpha$ 
radioactivity, where the regularity 
manifests itself as striking linear correlations between the logarithm of 
the decay half-life and the kinetic energy of the outgoing particle. 
This is known as the Geiger-Nuttall law, which was proposed 100 years 
ago \cite{gn}. This law  has been so successful that even today it is applied 
in studies of radioactive decays (see, e.g., Refs. \cite{Royer,Zag}). The reason for this success is
that the $\alpha$-particle formation probability, which is neglected 
in the Geiger-Nuttall law,  usually varies from
nucleus to nucleus much less than the penetrability. In
the logarithm scale of the Geiger-Nuttall law the differences in the
formation probabilities are usually small fluctuations along the straight
lines corresponding to the isotopic chains.
The importance of a proper treatment of $\alpha$ decay was attested by a
recent calculation which shows that the different lines can be merged in a
single one~\cite{qi08}.
The resulting universal decay law (UDL), which uses three free parameters only, 
explains well all known ground-state to ground-state radioactive decays. A similar three-term formula is also proposed in Refs. \cite{ren04,Ni08}.

This good agreement is a consequence of the smooth transition in the nuclear
structure that is often found when going from a nucleus to its neighboring
nuclei. This is also the reason why the BCS approximation works so
well in many nuclear regions. Notable discrepancies are only seen in a few 
cases around the $N=126$ shell closure where clustering induced by the pairing 
mode is inhibited \cite{Qi10}. Nuclei can even undergo more complex heavy-cluster decays, but these
also follow similar linear relations \cite{qi08,ren04,Ni08,Poe11,Poe11a}.

Besides these cluster decay cases, proton radioactivity also provides a unique
opportunity to study the structure of nuclei close to the proton
drip-line \cite{Hofmann89}. In the past decades tens of proton decay events 
have been observed in odd-$Z$ elements between $Z=53$ and $Z=83$, 
leading to an almost complete
identification of the edge of nuclear stability in this region
\cite{Woo97,Blank08}. On the theoretical side, one may extract from proton
emission detailed information on the 
nuclear potential beyond the drip-line \cite{Buck92,Ferreira02} as well as
the spectroscopic properties of the nuclei involved \cite{Aberg97,Davids96,Del06a}.

The proton-emission process can be looked as a quantum
tunneling through the Coulomb and centrifugal barriers
of a quasistationary state  \cite{Del06}. Similar to $\alpha$ and heavy 
cluster decays, the proton decay process can be divided into an ``internal 
region", where the compound state is restricted,
and the complementary ``external region". This division is such that in the
external region only the Coulomb and centrifugal forces are important and
the decaying system behaves like a two-particle system. The corresponding
expression for the half life can be written as \cite{Del10}
\begin{equation}\label{life}
T_{1/2}=\frac{\hbar\ln2}{\Gamma_l} = \frac{\ln2}{\nu} \left|
\frac{H_l^+(\chi,\rho)}{R \mathcal{F}_l(R)} \right|^2,
\end{equation}
where $\nu$ and $l$ are the outgoing velocity and the angular momentum carried by the outgoing proton, respectively.
$\mu$ is the reduced mass corresponding to the final system.
$R$ is the radius dividing the internal and external regions. The half life does not depend upon $R$
\cite{Del06}. At this
point the wave function of the proton is matched with the corresponding 
outgoing wave function 
in the external region.  In the following, the distance $R$  will be taken as the touching point, i.e.,
$R=R_0(A_d^{1/3}+1)$, with $R_0$=1.2 fm.
The other quantities are standard, i.e., $H_l^+(\chi,\rho)$ is the
Coulomb-Hankel function with arguments $\chi=2Z_d e^2/\hbar \nu$ and
$\rho=\mu\nu R/\hbar$. 

The amplitude of the wave
function in the internal region is the formation amplitude,
\begin{equation}\label{foram}
{\cal F}_l(R)=\int d{\mathbf R} d\xi_d 
[\Psi(\xi_d)\xi_pY_l(\mathbf R)]^*_{J_mM_m}
\Psi_m(\xi_d,\xi_p,\mathbf{R}),
\end{equation}
where $d$, $p$ and $m$ label the daughter, proton and mother
nuclei, respectively. $\Psi$ are the intrinsic wave functions and $\xi$ the corresponding
intrinsic coordinates. 
One sees from Eq.~(\ref{foram}) that $\mathcal{F}_l(R)$ would indeed be the
wave function of the outgoing particle $\psi_p(R)$ if the
mother nucleus would behave at the point $R$ as
\begin{equation}\label{mother}
\Psi_m(\xi_d,\xi_p,\mathbf{R})=
[\Psi(\xi_d)\xi_p\psi_p(R) Y_l(\mathbf R)]_{J_mM_m}.
\end{equation}

By applying a similar technique as in Ref. \cite{qi08} and considering the influence of the centrifugal barrier, one finds
that the logarithm of the decay half-life can be approximated by
\begin{eqnarray}
\label{gn-2} 
\log T_{1/2}  &=&a\chi' + b\rho' +d l(l+1)/\rho' + c,
\end{eqnarray}
where $a$, $b$, $c$ and $d$ are constants, $ \chi' =
A^{1/2}Z_pZ_dQ_p^{-1/2} $, $\rho' = \sqrt{AZ_{p}
Z_d(A_d^{1/3}+A_p^{1/3})}$, $A=A_dA_p/(A_d+A_p)$, and in this case of proton
decay it is $Z_p$ = $A_p$ =1 (for details see Ref. \cite{qi08}). Notice that 
$c$ depends on the formation amplitude $\mathcal{F}_l(R)$. This formula should be compared with that of Ref. \cite{Del06a} where a linear correlation between the reduced half-life and the Coulomb parameter was found.

The coefficients $a$ to $d$ can be determined by fitting available 
experimental data. For this purpose we took a total number of 44 decay events 
from the compilation of Ref. \cite{Blank08} and the recent results of Ref. 
\cite{Darby11}. The fitted values are listed in Table \ref{table1}. It is seen
that the log values of the experimental half-lives can be reproduced within 
an error of $\sigma=0.44$. Correspondingly, the decay half-lives are
reproduced within an average factor of three. This should be compared with the case of $\alpha$ decay where the UDL reproduces the available experimental half-lives within a factor of about 2.2.

\begin{table}
\centering \caption{Coefficient sets of Eq.~(\ref{gn-2}) that
determined by fitting to experiments of all proton decays (I) and to decays 
of nuclei with $N<75$ and $N\geq75$ separately (II), and the corresponding 
rms deviations.\label{table1}}
\begin{ruledtabular}
\begin{tabular}{ccccccc}
&Emitter& a& b&c &d & $\sigma$ \\
\hline
I&all&0.386&-0.502&2.386&-17.8&0.440\\
\multirow{2}{*}{II}&$N<75$&0.443&-0.364&2.66&-23.6&0.344\\
&$N\geq75$&0.403&-0.110&2.766&-27.8&0.235\\
\end{tabular}
\end{ruledtabular}
\end{table}

\begin{figure}
\includegraphics[width=0.45\textwidth]{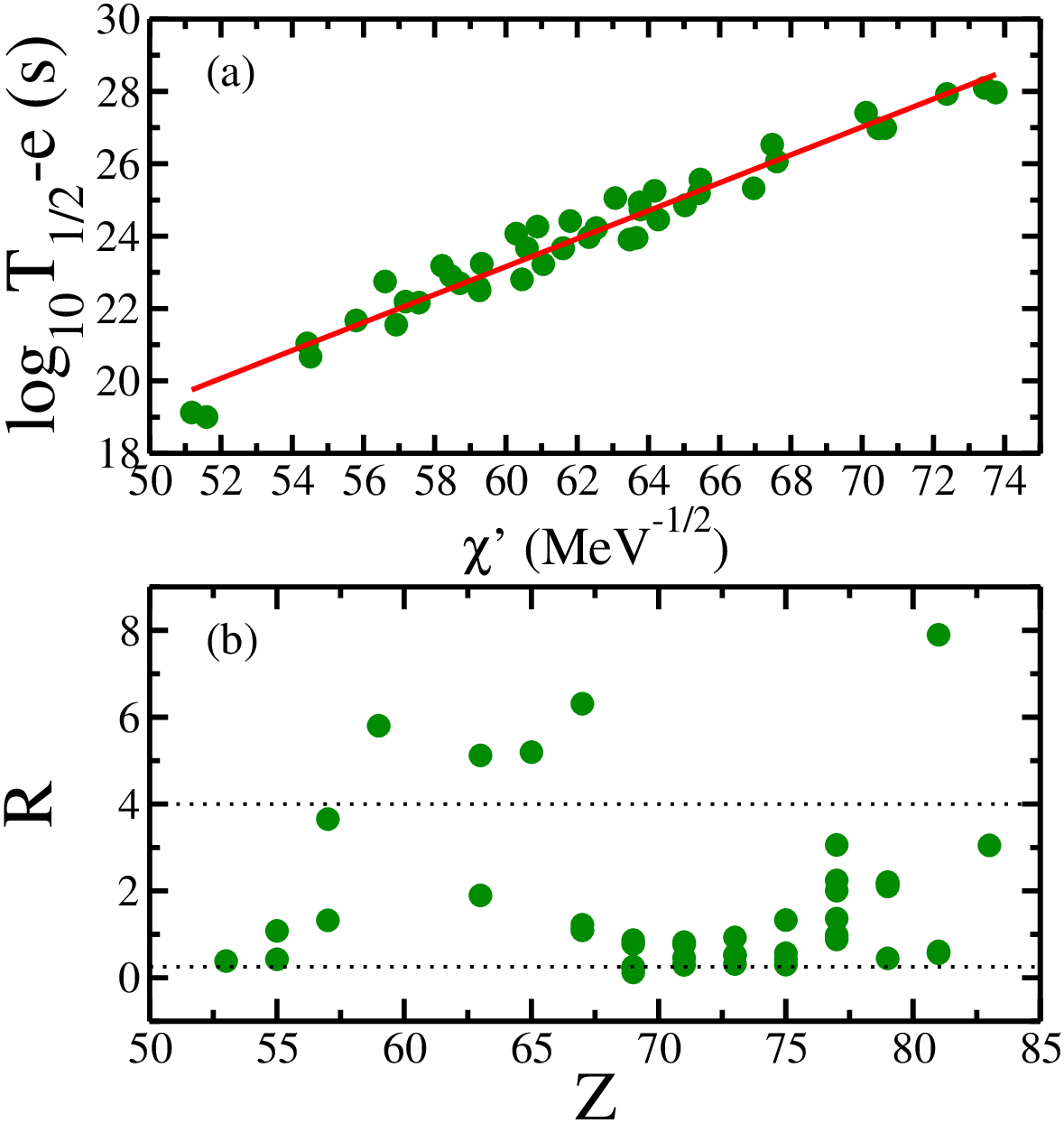}
\caption{(Color online) Upper: UDL proton decay half-lives. 
Dots are experimental values while the straight line is $a\chi'$ with $a$ 
provided by the parameter set I of Table I. $e$ is given by 
$e=b\rho' +d l(l+1)/\rho' + c$. 
Lower: ${\cal R}=T^{{\rm Expt.}}_{1/2}/T^{{\rm Cal.}}_{1/2} $ as a function of the
charge number $Z$. Within the dashed lines is the region where the
data is reproduced by the calculation within a factor of 4.  \label{fig1}}
\end{figure}

To probe Eq. (\ref{gn-2}), we plotted in the upper panel of  Fig. \ref{fig1} the quantity 
$ \log T^{{\rm Expt.}}_{1/2}- [b\rho' +d l(l+1)/\rho' + c]$ as a function of $\chi'$.
In the lower part of this Figure we plotted the discrepancy between
experimental and calculated values, i.e., the ratio
${\cal R}=T^{{\rm Expt.}}_{1/2}/T^{{\rm Cal.}}_{1/2} $, as a function of the 
emitter charge numbers $Z$. The calculations were performed by using
the parameter set I in Table \ref{table1}. 
It is seen that most of the data can be reproduced by the calculation within a 
factor of 4, i.e., with $0.25\leq {\cal R}\leq4$. Larger discrepancies are seen for a 
emitters between $63\leq Z\leq 67$ and the isomeric $h_{11/2}$
hole state in  the $Z=81$ nucleus $^{177}$Tl, where the experimental decay 
half life is underestimated by the calculation by a factor of about 8.

\begin{figure}
\includegraphics[width=0.45\textwidth]{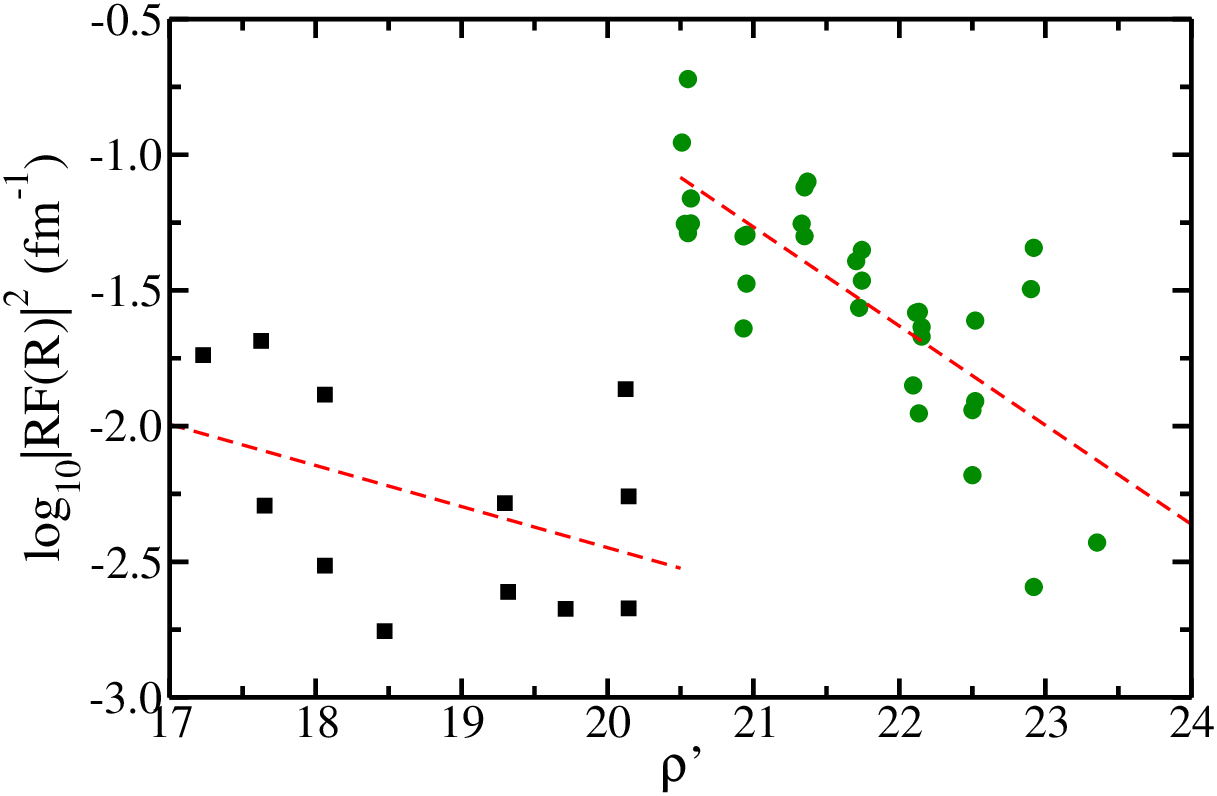}
\caption{(Color online) Proton-decay formation amplitudes 
$\log_{10} |RF(R)|^{2}$ extracted from experimental data as a function of 
$\rho'$.  Squares correspond to nuclei with $N<75(Z\leq67)$ while circles 
are for $N\geq 75(Z>67)$.
\label{fvsrho}}
\end{figure}

A better understanding of these discrepancies would require 
a systematic study  of the formation probabilities. We performed this task 
by extracting the formation amplitudes $F(R)$ from the experimental half lives 
by using the expression \cite{qi08},
\begin{equation}
\label{forpro}
\log |RF(R)|^{-2}=\log T^{{\rm Expt.}}_{1/2} - \log \left[ \frac{\ln
2}{\nu}|H^+_l(\chi,\rho)|^2\right],
\end{equation}
which we plotted in Fig.~\ref{fvsrho} as a function of $\rho'$. 

One sees in this Figure that two clearly defined regions emerged, as 
indicated by the dashed lines which we
obtained by using the  parameter set II of Table \ref{table1}.

The origin of these two regions can be gathered from the values of the $Q$ values,
deformation parameters and formation probabilities 
presented in Table \ref{table2}, where
the experimental and corresponding theoretical half-lives are also given.

\begin{table*}
\caption{Experimental and calculated half-lives for proton decays to
ground states. Experimental $Q$ values and half-lives are taken from Refs.
\cite{Blank08,Darby11}. Stars in the emitters indicate isomeric states. 
The deformation parameter
$\beta_2$ are from Ref. \cite{Moller95}. The calculations were performed
by using the parameter sets I ($T_{1/2}^{\text {Cal.}}$(I))
and II ($T_{1/2}^{\text {Cal.}}$(II)) of Table \ref{table1}. In the last two 
columns are the formation amplitudes $|F_l(R)|$ and formation
probabilities $|RF_l(R)|^2$ (Eq. (\ref{forpro})). Units in the calculated
half lives are as in the corresponding experimental values.
\label{table2}}
\begin{ruledtabular}
\begin{tabular}{cccccccccccc}
Emitter & ~~$l$~~ & ~~$j_m^\pi$~~ & $Q (\text{keV}) $&
$\beta_2$& $T_{1/2}^{\text {Expt.}}$  &$T_{1/2}^{\text {Cal.}}$(I)&
$T_{1/2}^{\text {Cal.}}$(II)&$|F_l(R)|$(fm$^{-3/2}$)& $|RF_l(R)|^2$(fm$^{-1}$) \\
\hline
$^{109}_{~53}$I~$^{~}$ & 2&~5/2$^{+}$& ~827(~5)& ~0.160&
93.5(5) $\mu$s& 243 & 187 &0.020&          0.018\\

$^{112}_{~55}$Cs$^{~}$ & (2)&(0$^{+},3^{+}$)& ~823(~7)& ~0.208&
0.5(1) ms&1.18 & 1.39 &0.021&          0.021\\

$^{113}_{~55}$Cs$^{~}$ & 2&~3/2$^{+}$& ~976(~3)& ~0.207&
16.7(7) $\mu$s&15.5 &9.72 &0.010&          0.005\\

$^{117}_{~57}$La$^{~}$ & 2&~3/2$^{+}$& ~814(11)& ~0.290&
23.8(20) ms&6.52 &12.3 &0.008&          0.003\\

$^{117}_{~57}$La$^{~}$ & 4&~9/2$^{+}$& ~951(~6)& ~0.290&
10(5) ms&7.55 &12.6 &0.016&          0.013\\

$^{121}_{~59}$Pr$^{~}$ & 2&~3/2$^{+}$& ~900(10)& ~0.318&
10($^{+6}_{-3}$) ms&1.72 &3.28 &0.006&          0.002\\

$^{130}_{~63}$Eu$^{~}$ & 2&~1$^{+}$& 1039(15)& ~0.331&
0.90($^{+49}_{-29}$) ms&0.474& 1.12 &0.010&          0.005
\\

$^{131}_{~63}$Eu$^{~}$ & 2&~3/2$^{+}$& ~959(~9)& ~0.331&
21.4($^{+1.8}_{-1.7}$) ms&4.18 &13.8 &0.007&          0.002\\

$^{135}_{~65}$Tb$^{~}$ & 3&~7/2$^{-}$& 1200(~7)& ~0.325&
0.94($^{+0.33}_{-0.22}$) ms&0.181 &0.428 &0.006&          0.002
\\

$^{140}_{~67}$Ho$^{~}$ & 3& $(6^{-},~0^-,~8^+)$& 1106(10)& ~0.297&
6(3) ms&4.91 &23.3 &0.016&          0.014\\

$^{141}_{~67}$Ho$^{~}$ & 3&~7/2$^{-}$& 1190(~8)& ~0.286&
4.1(1) ms&0.650 &2.30 &0.006&          0.002\\

$^{141}_{~67}$Ho$^{*}$ & 0&~1/2$^{+}$& 1255(~8)& ~0.286&
6.6($^{+0.9}_{-0.7}$) $\mu$s&6.04 &12.0 &0.010&          0.006\\

$^{144}_{~69}$Tm$^{~}$ & 5& (10$^+$, 5$^-$)& 1725(16)& ~0.258&
2.7($^{+1.7}_{-0.7}$) $\mu$s&22.0 &6.37&0.045&          0.111\\

$^{145}_{~69}$Tm$^{~}$ & 5&11/2$^{-}$& 1753(~7)& ~0.249&
$\sim$3.46(32) $\mu$s&14.8 &4.28 &0.031&          0.056\\

$^{146}_{~69}$Tm$^{~}$ & 5&(5$^-$)& 1210(~4)& -0.199&
117.6(64) ms & 151 &67.8 &0.030&          0.051\\

$^{146}_{~69}$Tm$^{*}$ & 5&(10$^+$)& 1140(~4)& -0.199&
203(6) ms&792 &383 &0.058&          0.190\\

$^{147}_{~69}$Tm$^{~}$ & 5&11/2$^{-}$& 1073(~5)& -0.190&
~3.78(1.27) s& 4.36 &2.32 &0.035&          0.069 \\

$^{147}_{~69}$Tm$^{*}$ & 2&~3/2$^{+}$& 1133(~3)& -0.190&
0.360(36) ms& 1.51 &0.269 &0.031&          0.056\\

$^{150}_{~71}$Lu$^{~}$ & 5&$>5^{-}$& 1283(~3)& -0.164&
64.0(56) ms&83.1 &51.3 &0.030&          0.050\\

$^{150}_{~71}$Lu$^{*}$ & 2&(1$^{-},2^-$)& 1306(~5)& -0.164&
43($^{+7}_{-5}$) $\mu$s& 94.1 &20.9 &0.020&          0.023\\

$^{151}_{~71}$Lu$^{~}$ & 5&11/2$^{-}$& 1253(~3)& -0.156&
127.1(18) ms& 155 &100 &0.030&          0.051\\

$^{151}_{~71}$Lu$^{*}$ & 2&~3/2$^{+}$& 1332(10)& -0.156&
16(1) $\mu$s&54.0 &11.9 &0.024&          0.033\\

$^{155}_{~73}$Ta$^{~}$ & 5&11/2$^{-}$& 1468(15)& ~0.008&
2.9($^{+1.5}_{-1.1}$) ms&5.56 &4.37 &0.031&          0.056\\

$^{156}_{~73}$Ta$^{~}$ & 2&(2$^-$)& 1032(~5)& -0.050&
149(8) ms &286 &135 &0.029&          0.050\\

$^{156}_{~73}$Ta$^{*}$ & 5&(9$^+$)& 1127(~7)& -0.050&
8.52(2.12) s&9.19 &10.3 &0.036&          0.076\\

$^{157}_{~73}$Ta$^{~}$ & 0&~1/2$^{+}$& ~947(~7)& ~0.045&
0.300(105) s&0.941 &0.401 &0.037&          0.080\\

$^{159}_{~75}$Re$^{*}$ & 5&11/2$^{-}$& 1831(20)& ~0.053&
20.2(37) $\mu$s&47.7 &42.6 &0.026&          0.041
\\

$^{160}_{~75}$Re$^{~}$ & 2&(2$^{-}$)& 1287(~6)& ~0.080&
0.687(11) ms &1.22 &0.642 &0.021&          0.027\\

$^{161}_{~75}$Re$^{~}$ & 0&~1/2$^{+}$& 1214(~6)& ~0.080&
0.440(2) ms&1.45 &0.660 &0.027&          0.045\\

$^{161}_{~75}$Re$^{*}$ & 5&11/2$^{-}$& 1338(~6)& ~0.080&
224(31) ms&169 &226 & 0.024&          0.034 \\

$^{164}_{~77}$Ir$^{~}$ & 5&(9$^{+}$)& 1844(~9)& ~0.089&
0.113($^{+62}_{-30}$) ms&0.0829 &0.108 &0.015&          0.014\\

$^{165}_{~77}$Ir$^{*}$ & 5&11/2$^{-}$& 1733(~7)& ~0.099&
0.34(7) ms&0.385 &0.550 &0.021&          0.026\\

$^{166}_{~77}$Ir$^{~}$ & 2& & 1168(~7)& ~0.107&
0.152(71) s&0.0678 &0.0631 &0.014&          0.011\\

$^{166}_{~77}$Ir$^{*}$ & 5&(9$^{+}$)& 1340(~8)& ~0.107&
0.84(28) s&0.419 &0.835 &0.021&          0.026\\

$^{167}_{~77}$Ir$^{~}$ & 0&~1/2$^{+}$& 1096(~6)& ~0.116&
110(15) ms&112 &0.919 & 0.020&          0.023\\

$^{167}_{~77}$Ir$^{*}$ & 5&11/2$^{-}$& 1261(~7)& ~0.116&
7.5(24) s&2.45 &5.39 & 0.019&          0.021\\

$^{170}_{~79}$Au$^{~}$ & 2& (2$^{-}$)& 1488(12)&~0.080 &
321($^{+67}_{-58}$) $\mu$s &153 &1.53 &0.010&          0.007\\

$^{170}_{~79}$Au$^{*}$ & 5& (9$^{+}$)& 1770(~6)&~0.080 &
1.046($^{+136}_{-126}$) ms &0.482 &0.991 & 0.014&          0.011\\

$^{171}_{~79}$Au$^{~}$ & 0&~1/2$^{+}$& 1464(10)& -0.105&
24.5($^{+4.7}_{-3.1}$) $\mu$s&54.8 &45.1 &0.020&          0.025\\

$^{171}_{~79}$Au$^{*}$ & 5&11/2$^{-}$& 1719(~4)& -0.105&
2.22(19) ms&1.01 &2.18 &0.014&          0.012\\

$^{176}_{~81}$Tl$^{~}$ & 0&(3$^{-}$, 4$^-$, 5$^{-}$)& 1282(18)&
-0.053&5.2($^{+3.0}_{-1.4}$) ms &8.55 &12.7 &0.023&          0.032  \\

$^{177}_{~81}$Tl$^{~}$ & 0&~1/2$^{+}$& 1180(20)& -0.053&
67(37) ms&119 & 203 &0.027&          0.045 \\

$^{177}_{~81}$Tl$^{*}$ & 5&11/2$^{-}$& 1984(~8)& -0.053&
396($^{+87}_{-77}$) $\mu$s &50.1 &137 &0.006&          0.003\\

$^{185}_{~83}$Bi$^{~}$ & 0&~1/2$^{+}$& 1624(16)& -0.052&
58(9) $\mu$s& 19.0 &33.1 &0.008&          0.004
\\
\end{tabular}
\end{ruledtabular}
\end{table*}

One sees in this Table that the region to the left in Fig. \ref{fvsrho},
i.e., for lighter isotopes, corresponds to the decays of well deformed nuclei. The formation probabilities decreases for these
nuclei as $\rho'$ increases. Then, suddenly, a strong
transition occurs at $\rho'$=20.5, corresponding to the nucleus
$^{144}_{~69}$Tm. Here the formation probability acquires its maximum value,
and then decreases again as $\rho'$ increases. One notices in Table 
\ref{table2} that this tendency is followed by the deformations, which
gradually diminishes after the transition point.  
One thus sees that the reason of this tendency of the formation amplitude is
related to the deformation. Indeed, in the left region of Fig. \ref{fvsrho},  the decays of the
deformed nuclei proceed through  
small spherical components of the corresponding deformed orbitals and,
therefore, the formation probabilities are small.

The right region of Fig. \ref{fvsrho} involves the decays of spherical orbits
as well as major spherical components of deformed orbitals (for example, $h_{11/2}$ component of the orbital $11/2 [505]$). 

Another striking feature is that 
the formation  probability of the odd-odd nucleus $^{144}$Tm is abnormally large.
Considering that the experimental uncertainties regarding the half-life
(from where the formation probability is extracted) is large one may doubt the 
validity of this anomaly. Moreover, it is  not surprising, given this anomaly,  that for $^{144}$Tm the half-life
evaluated with the set of parameters I differs from the corresponding 
experimental value by a factor $R$
which is beyond the scale used in the lower panel of Fig. \ref{fig1}.  

\begin{figure}
\includegraphics[width=0.45\textwidth]{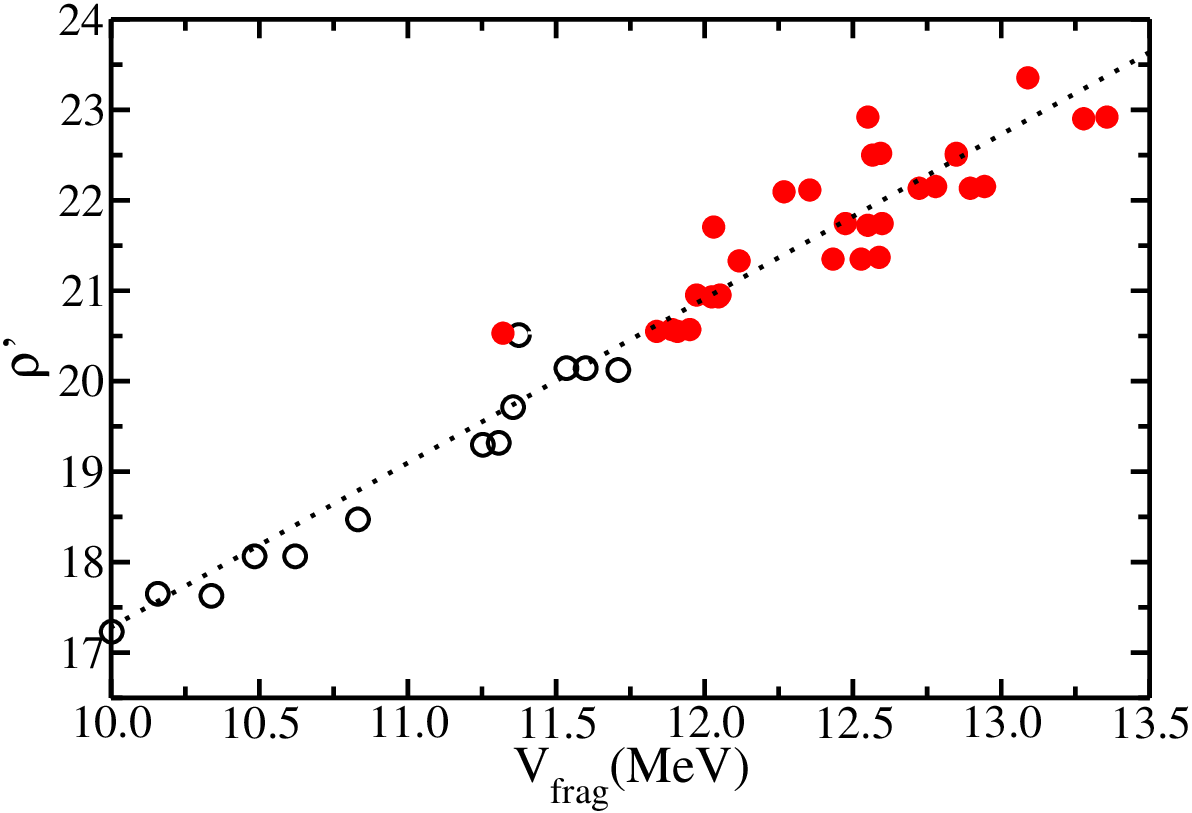}
\caption{The parameter $\rho'$ versus the fragmentation
potential (Eq. (\ref{Vfrag})) for nuclei with $Z\leq67$ (open circles)
and $Z>67$ (dark circles).
\label{rhovsfrag}}
\end{figure}

It is interesting to point out that a similar ``clustering" of the proton
wave functions into two distinct regions, as the one seen in Fig. \ref{fvsrho},
was  evidenced in Ref. \cite{Del09}. But there the determining
variable was not $\rho'$ but the so-called fragmentation
potential. This is given by the difference between the Coulomb barrier and the
$Q$-value as
\begin{equation}
\label{Vfrag}
V_{{\rm frag}}=\frac{Z_d e^2}{R}-Q.
\end{equation}
This feature is not surprising, since there is a linear correlation between
these parameters, as can be seen in Fig. \ref{rhovsfrag} where
the parameter $\rho'$ is plotted versus the above defined fragmentation
potential. One sees here a clear linear correlation between these
parameters. Moreover, the plot is divided into two regions
with $Z\leq67$ and $Z>67$.
In fact that correlation is stronger than what the figure suggests since the upper part of the plot (dark circles) is divided into
several regions where few close values of the fragmentation potential (with different $Q$ values) correspond
to the same $\rho'$. 
It is worthwhile to point out that the linear correlation  between $V_{{\rm frag}}$ and $\rho'$, as seen in Fig. \ref{rhovsfrag}, is a result of the matching of the proton internal wave function with the corresponding external Coulomb wave. This is because both parameters are connected by this procedure.

\begin{figure}
\includegraphics[width=0.45\textwidth]{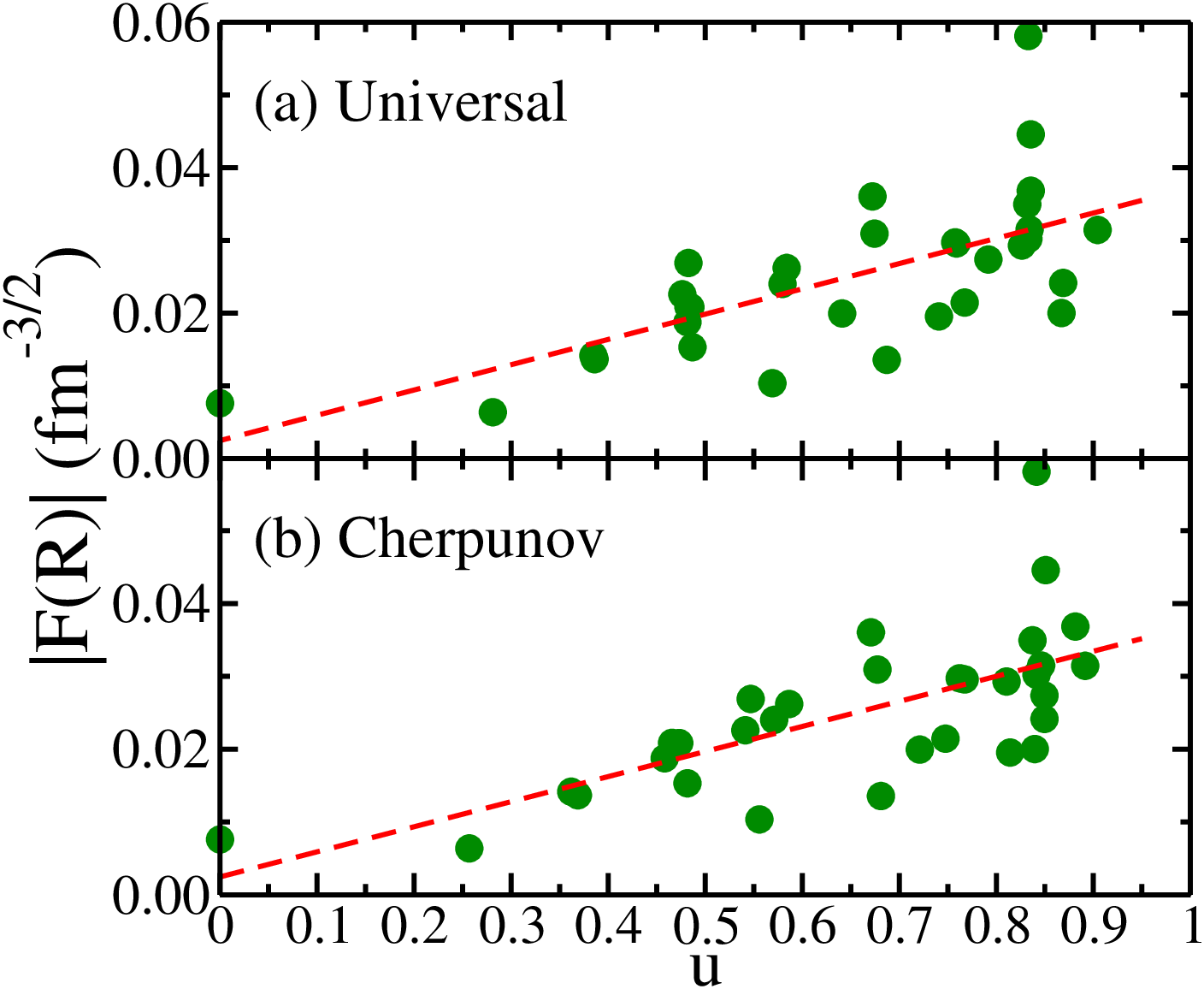}
\caption{(Color online)
The formation amplitudes $|F_l(R)|$ extracted from experimental data for proton
decays of nuclei  $N\geq 75(Z>67)$ as a function of $u$ calculated from 
BCS calculations using for the Woods-Saxon mean field the universal parameters 
\cite{uni} (upper) and the Cherpunov parameters \cite{che67} (lower). 
\label{fvsu}}
\end{figure}

Going back to Fig. \ref{fvsrho} we notice that in the BCS approach the formation amplitude at a given radius $R$ would be 
proportional to the product of the occupancy $u$ times the single-proton wave 
function $\psi_p(R)$. Therefore the tendencies seen in Fig. \ref{fvsrho} may
be due to the BCS amplitudes or the radial wave functions. 
To recognize the influence of the value of the BCS amplitudes  we plotted 
in Fig. \ref{fvsu} the formation probabilities $|F_l(R)|$ extracted 
from experiment for the case of proton
decays corresponding to nuclei with $N\geq 75(Z>67)$ as a function of $u$. 
The $u$ values were calculated by using a Woods-Saxon potential according to 
the  universal (upper) and Cherpunov (lower) parameters. One sees that the
tendency is the same for both sets of parameters, namely that the values of
of the formation probabilities increases with $u$, strengthening our
interpretation of the behavior of the formation probabilities presented
above.  

\begin{figure}
\includegraphics[width=0.45\textwidth]{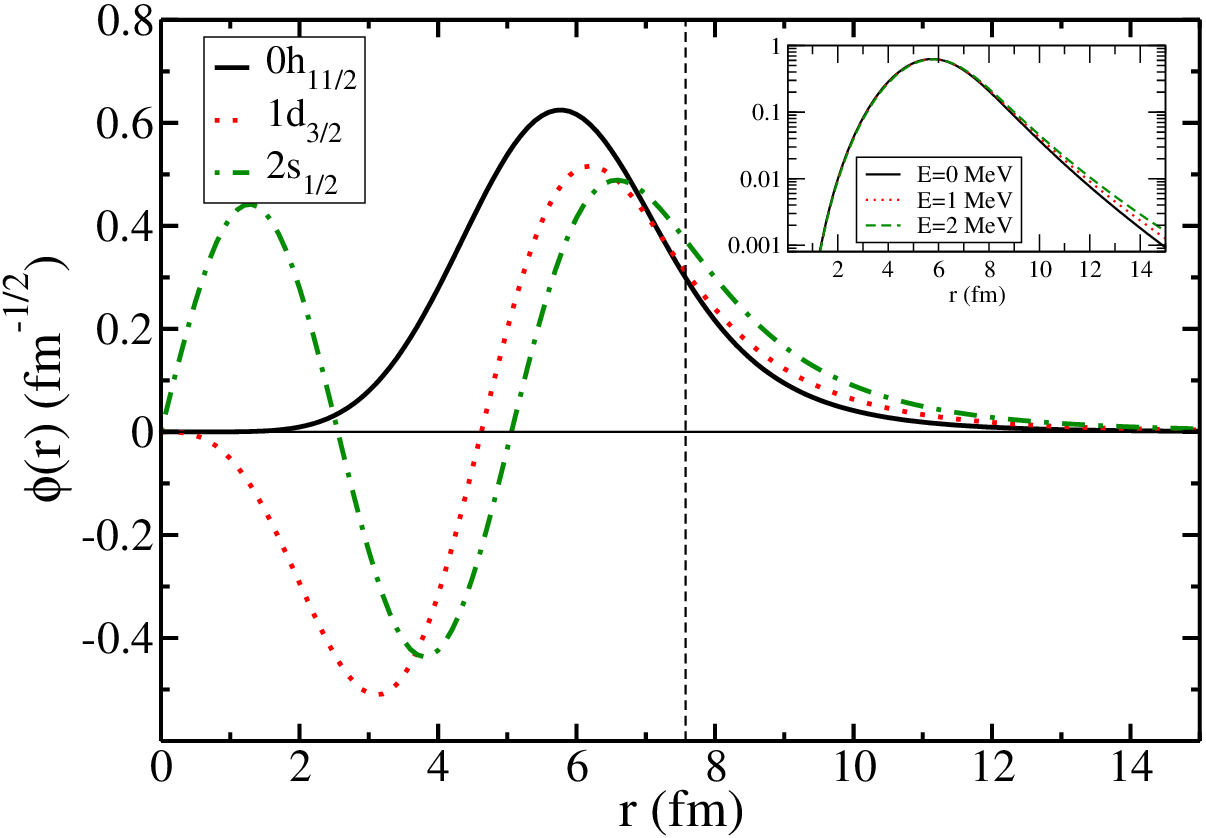}
\caption{(Color online) Single-proton wave functions in $^{151}$Lu for 
different channels calculated with the universal Woods-Saxon parameter. The 
inserted plot shows the wave functions of the $0h_{11/2}$ orbital with 
different energies derived by changing the depth of the potential. 
\label{single}}
\end{figure}

Still one has to consider eventual changes in the radial wave functions at
the nuclear radius (where the formation probabilities are determined). For
this we plotted in Fig. \ref{single} the single-particle wave functions $\phi(r)=r\psi_p(r)$ 
for different orbitals corresponding to the nucleus $^{151}$Lu. It is 
seen that the values of $\phi(r)$ around the nuclear surface are quite similar 
to each other. From this we conclude that the fluctuations in the experimental 
formation amplitudes found above
for nuclei with $N\geq75$ are mainly due to fluctuations of the $u$ values.
It is also to be noticed that some cases which departs from the UDL
correspond to the decays of hole states. This occurs in the isomeric state of 
$^{177}$Tl (as already pointed out above) and also in the ground state 
of $^{185}$Bi  \cite{Poli99}.

Another feature that has to be considered is the influence (if any) of the  $Q$
value upon the formation probabilities. The
$Q$ value determines the penetrability and, therefore, the radioactive decay
process. The question is whether even the spectroscopic quantities are
affected by the $Q$ value. 
In Ref. \cite{Del09} it was evidenced the fact that the logarithm of the 
wave function is  proportional to the fragmentation potential (Eq. (\ref{Vfrag}))
for all decay processes, in particular for proton emission. This is a direct
consequence of the one-body Schr\"odinger equation. Anyway, it seems that
the dependence on the $Q$-value alone is small in proton decay processes.
To analyze this we show  in the inserted plot of Fig. \ref{single}
the wave  functions of the $0h_{11/2}$ orbital under different energies. These were
obtained by changing the depth of the potential. As perhaps expected, 
the formation amplitude at the nuclear surface is not sensitive to changes
in the energy, i.e., in the $Q$ values. Neither the amplitudes $u$ are much 
affected by the changing of the potential depth, as also expected.
One therefore concludes that the changes in the formation amplitudes
extracted from experimental data are a result of changes in the deformation
of the decaying nuclei.


In conclusion, we have shown in this paper that the proton formation probability in the
corresponding decay process depends upon the deformation of the decaying
nucleus. In a well deformed nucleus the decay proceeds through one of the
spherical components of the deformed orbit, which is usually small in this
case of large deformations. Therefore the formation probability is small. On
the contrary, in spherical or weakly deformed nuclei the decay proceeds
through the only component that is available and, as a result, the formation
probability is large. We have strengthen this interpretation by showing that
in the case of weakly deformed nuclei the formation probability is
proportional to the BCS
occupation number $u$ corresponding to the decaying orbit.

This work was supported by the Swedish Research Council (VR) under grant Nos. 623-2009-7340 and 621-2010-4723
and by a grant of the Romanian National Authority for Scientific Research, CNCS-UEFISCDI, project number PN-II-ID-PCE-2011-3-0092.
CQ also acknowledges the computational support provided by the Swedish National Infrastructure for Computing (SNIC)
at PDC and NSC.

\end{document}